\documentclass[aps,twocolumn,prl,preprintnumbers,amsmath,amssymb,superscriptaddress]{revtex4}

\usepackage{graphicx}
\usepackage{bm}
\usepackage{hyperref}

\begin{document}

\title{Low energy electrodynamics of the Kondo-lattice antiferromagnet CeCu$_2$Ge$_2$}

\author{ G. Boss\'e}
\affiliation{The Institute for Quantum Matter, Department of Physics and Astronomy, The Johns Hopkins University, Baltimore, MD 21218 USA.}
\author{L. S. Bilbro}
\affiliation{The Institute for Quantum Matter, Department of Physics and Astronomy, The Johns Hopkins University, Baltimore, MD 21218 USA.}
 \author{R. Vald\'es Aguilar}
\affiliation{The Institute for Quantum Matter, Department of Physics and Astronomy, The Johns Hopkins University, Baltimore, MD 21218 USA.}
 \author{LiDong Pan}
\affiliation{The Institute for Quantum Matter, Department of Physics and Astronomy, The Johns Hopkins University, Baltimore, MD 21218 USA.}
 \author{Wei Liu}
\affiliation{The Institute for Quantum Matter, Department of Physics and Astronomy, The Johns Hopkins University, Baltimore, MD 21218 USA.}
\author{A. V. Stier}
\affiliation{The Institute for Quantum Matter, Department of Physics and Astronomy, The Johns Hopkins University, Baltimore, MD 21218 USA.}
\author{Y. Li}
\affiliation{Department of Physics and Fredrick Seitz Materials Research Laboratory, University of Illinois at Urbana-Champaign, Urbana, Illinois, 61801 USA.}
\author{J. Eckstein}
\affiliation{Department of Physics and Fredrick Seitz Materials Research Laboratory, University of Illinois at Urbana-Champaign, Urbana, Illinois, 61801 USA.}
 \author{N. P. Armitage}
 \affiliation{The Institute for Quantum Matter, Department of Physics and Astronomy, The Johns Hopkins University, Baltimore, MD 21218 USA.}

\date{\today}

\begin{abstract}

We present time-domain THz spectroscopy data of a thin film of the Kondo-lattice antiferromagnet CeCu$_2$Ge$_2$. The low frequency complex conductivity has been obtained down to temperatures below the onset of magnetic order. At low temperatures a narrow Drude-like peak forms, which is similar to ones found in other heavy fermion compounds that do not exhibit magnetic order. Using this data in conjunction with DC resistivity measurements, we obtain the frequency dependence of the scattering rate and effective mass through an extended Drude model analysis.  The zero frequency limit of this analysis yields evidence for large mass renormalization even in the magnetic state, the scale of which agrees closely with that obtained from thermodynamic measurements.

\end{abstract}



\maketitle

Kondo lattice systems in which localized $f$ moments hybridize with extended conduction band electrons exhibit a rich variety of behavior.  At high temperatures these systems behave as an ensemble of weakly interacting conduction electrons with modest masses and free $f$-moments.   At low temperatures,  the hybridization of local moments with the conduction electrons results in phenomena such as metallic states with charge carriers whose mass is hundreds of times the free electron mass (so called ``heavy-fermion" behavior), superconductivity, and various interesting magnetic states \cite{Fisk1988a,Pfleiderer09a,Steglich91a}.    Which physics is expressed at low temperature is determined by a delicate balance and competition between the Ruderman-Kittel-Kasuya-Yosida (RKKY) type magnetic interaction and the hybridization of the $4f$ or $5f$ electrons with delocalized states \cite{Doniach77a}.  Weak hybridization typically yields the RKKY-type interaction and magnetically ordered ground states.   Strong hybridization, on the other hand, leads to the formation of fluctuating valence states, compensation of local moments, and heavy-electron metals.  In the limiting case of very strong hybridization, the delocalized band electrons screen the localized $f$ electrons, forming a singlet.  Kondo lattice systems typically exhibit their heavy-electron effects below a temperature scale $T^*$ at which successive scattering from local moments starts to be in-phase and collective effects can appear.

The competition between Kondo screening and the RKKY interaction is exhibited dramatically in the evolution of the electronic structure by Si substitution in the CeCu$_2$Ge$_{2-x}$Si$_x$ Kondo-lattice series or by the application of pressure to the pure compound.  Due to its smaller ionic size, Si substitution can be seen as a equivalent to increasing pressure.  Pure CeCu$_2$Ge$_{2}$ (CCG) is a Kondo-lattice system that exhibits an unusual temperature dependence of the resistivity caused by crystal field splitting and magnetic interactions at low temperatures  \cite{Felten87a,Knopp89a,Yi2011a}.   An upturn of the resistivity around 25 K indicates the increasing effect upon cooling of scattering of the conduction electrons off the local moments.   At approximately a temperature $T^* \sim$ 6 K, the resistivity reaches a maximum, giving evidence for a low Kondo lattice temperature scale, before ordering antiferromagnetically just slightly below $T^*$ at a N\'eel temperature $T_N$ = 4.15 K.  With the substitution of Si or increased pressure, the Kondo lattice temperature scale increases at the expense of the RKKY scale and the magnetic state is eventually suppressed \cite{Knebel96a,Jaccard1992a}.   At stoichiometries approaching $x=2$ or pressures above 70 kbar the Kondo physics dominates dramatically with the emergence of superconductivity out of a heavy electron state \cite{Jaccard1992a,Knebel96a}.   A regime of superconductivity mediated by magnetic fluctuations may give over at even higher pressures to a regime mediated by valence fluctuations.

In the pure CCG compound, the energy scales of the RKKY and the Kondo-type interactions are of the same order of magnitude \cite{Knopp89a} as demonstrated by the fact that the Kondo temperature only slightly exceeds $T_N$.  In the antiferromagnetic state the moments are partially Kondo compensated from 1.5 to 0.7 $\mu_B$.  The local moment antiferromagnetic order is sinusoidally modulated and incommensurate with a wavevector $\vec{q} \approx (0.284, 0.284,0.543)$ \cite{Knopp89a,Krimmel97a}.  Although many aspects of the physics fit the conventional model of RKKY and Kondo screening competition, there are a number of outstanding issues.  The large linear term of the electronic specific heat at temperatures well below the magnetic transition suggests that long-range magnetic order may coexist with a coherent heavy-fermion state \cite{deBoer87a}.   This is in contrast to the conventional view that the occurrence of the magnetic state should suppress the Kondo effect  \cite{Doniach77a}.  The low temperature electronic specific heat coefficient in CCG, $\gamma_e$, in the magnetic state is reported to be about 100  mJ/K$^2$ mol, which indicates effective masses of about 50-100 times the free electron value  \cite{deBoer87a}.

In this work we use time-domain THz spectroscopy (TDTS) to study the low frequency complex conductivity of thin films of CCG.  As in other recent papers \cite{Degiorgi99a,Dressel02a,Scheffler05a,Guritanu08a} that measure low frequency conductivity in heavy fermion compounds, we find evidence of mass enhancement through a dynamic measurement.  At temperatures well above the onset of magnetic order and the coherence temperature, the Drude conductivity is broad and featureless, which is consistent with the strong scattering in the normal state.  At low temperatures a narrow Drude-like peak develops at $\omega = 0$.  An extended Drude model analysis reveals that this can be connected to a strongly temperature and frequency dependent renormalization of the mass and scattering rate that develops at low temperature.  This mass renormalization in the low frequency and temperature limit is of order 50-100 times the band mass m$_b$, which is consistent with the enhancement inferred from specific heat measurements \cite{deBoer87a}.   Interestingly, these mass enhancements are found at temperatures below $T_N$ demonstrating that the development of magnetism does not quench the Kondo interaction.

\begin{figure}
\begin{center}
\includegraphics[width=1\columnwidth]{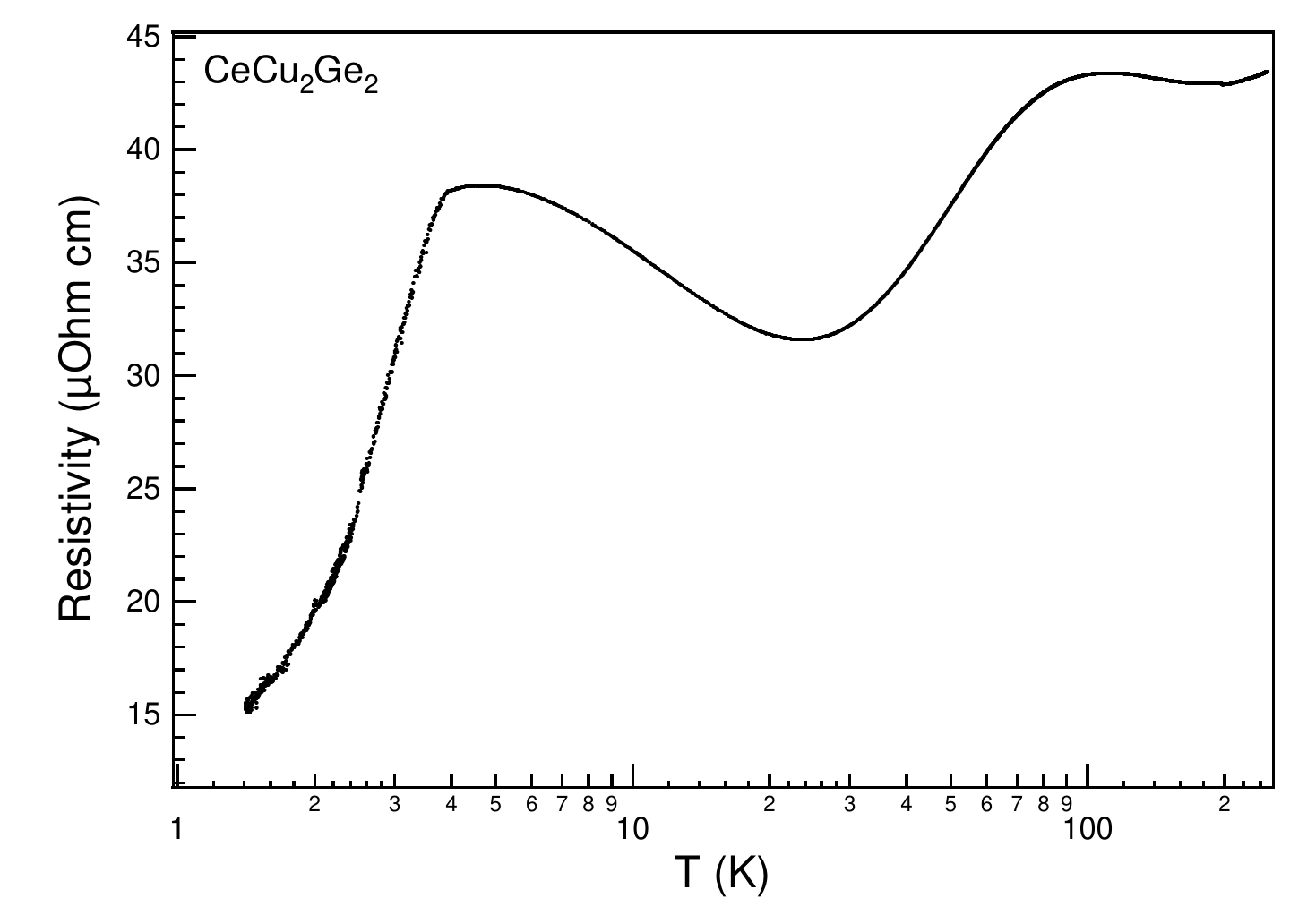}
\centering
\caption{Resistivity as a function of temperature for CCG film. Around 110K there is a broad peak corresponding to crystal field splitting; at 5.5K there is a sharp peak attributed to the Kondo lattice coherence temperature, $T^*$; a sharp cusp is seen near 4K signaling the onset of antiferromagnetic order.}
\label{Resist}
\end{center}
\end{figure}

In the technique of TDTS an infrared femtosecond laser pulse is split between two paths and excites a pair of ``Auston" switch photoconductive antennae; one acts as an emitter and the other as a receiver.  When the laser pulse hits the voltage biased emitter a broadband THz pulse is produced and collimated by mirrors and lenses and passes through the sample.  The THz pulse then falls on the receiving Auston switch.  Current only flows across the receiver switch at the instant the other short femtosecond pulse impinges on it. By varying the difference in path length between the two pulses, the entire electric field of the transmitted pulse can be mapped out as a function of time.  By dividing the Fourier transform of transmission through the sample by the Fourier transform of transmission through a reference substrate, one obtains the full complex transmission $T(\omega)$ over a frequency range that can be as broad as 100 GHz - 3.5 THz.  The complex transmission is then used to calculate the complex conductivity $\sigma$ without the need of Kramers-Kronig transformation using the expression $T(\omega)= \frac{1 + n }{1+n + \sigma d Z_0}$.  Here $n$ is the index of refraction of the substrate, $d$ is the film thickness, and $Z_0$ is the impedance of free space (377 Ohms).  The films studied in this work were grown by molecular beam epitaxy (MBE) using flux matched codeposition of the constituent atoms on MgO substrates.  In- situ RHEED and ex-situ AFM and XRD were used post-growth to characterize the films. Details of the growth have been reported elsewhere \cite{Yi2011a}.

In Fig. \ref{Resist} we show DC resistivity data as a function of temperature of the film studied in this work.   As demonstrated previously, transport measurements of the films find quantitative agreement with the temperature scales found in bulk single crystals  \cite{Yi2011a}.  A number of prominent features can be noted.   The broad peak at 110K has been attributed  to crystal field splitting.  The peak at 5.5 K has been attributed to the Kondo lattice coherence temperature, $T^*$, below which heavy quasi-particles are believed to form.  At slightly lower temperatures (4K), the antiferromagnetic transition occurs and the resistivity drops rapidly.

\begin{figure}
\begin{center}
\includegraphics[width=1\columnwidth]{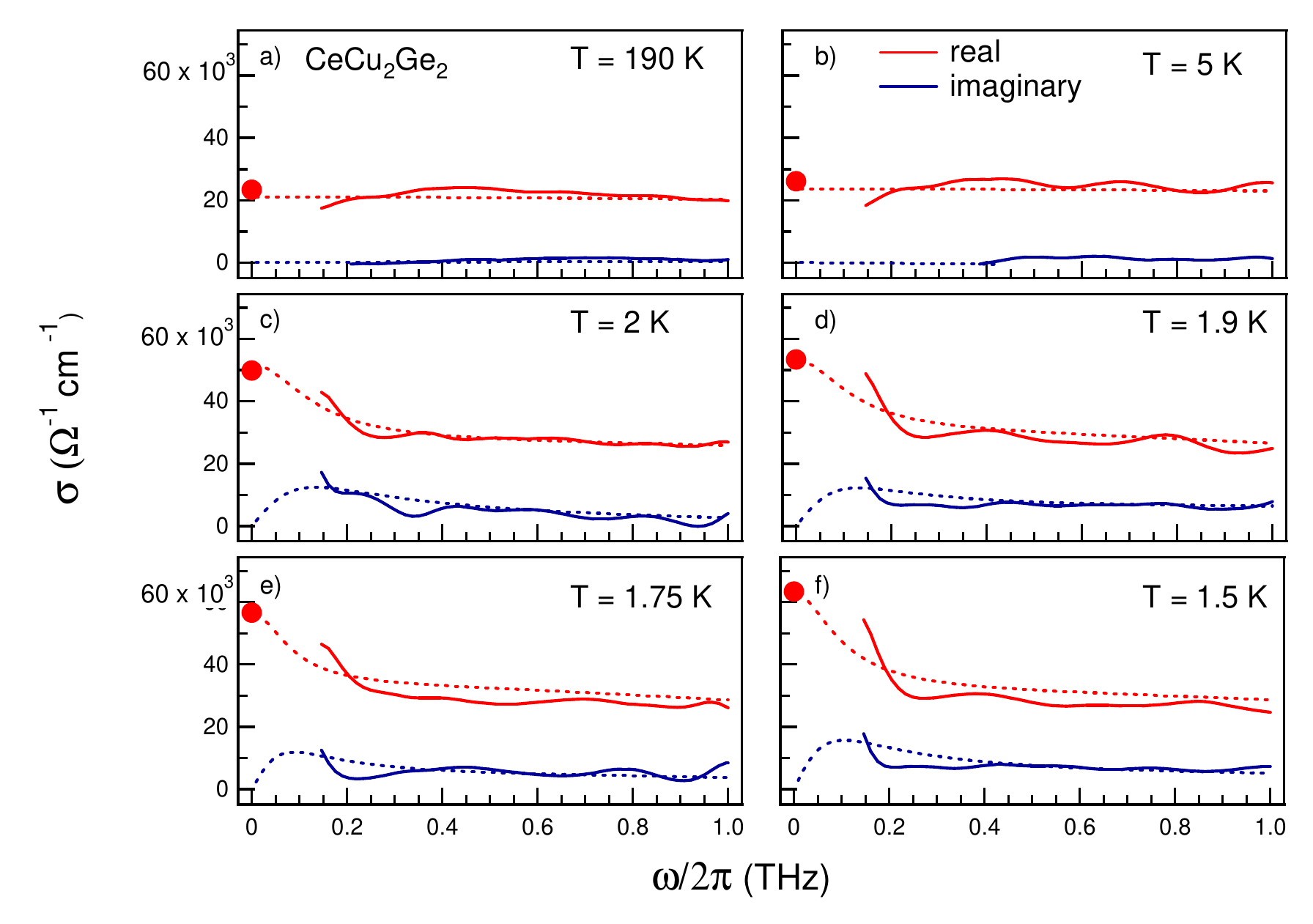}
\centering
\caption{(Color Online) a-f) Real and imaginary parts of the complex conductivity as a function of frequency for six temperatures in the frequency range of 0-1 THz.  Solid lines indicate experimental data, while dashed lines show results of a fit described in the text.  The values of the DC conductivity are also shown as a solid circle.}
\label{Cond}
\end{center}
\end{figure}

In Fig. \ref{Cond}a-f we show the real and imaginary parts of the complex conductivity $\sigma(\omega)$ for six different temperatures as a function of frequency.  Experimental data is shown from 150 GHz to 1 THz.  The comparatively reduced spectral range comes from the relative high conductivity of these films and consequent reduction in transmission.  The corresponding conductivities at zero frequency (from the DC resistance data) for each temperature are also shown.  At high temperatures the conductivity curves are broad and featureless reflecting the strong scattering of charge at temperatures well above $T^*$.  As expected from the resistivity data,  as the temperature is lowered the low-frequency conductivity becomes strongly temperature dependent.  Using a combination of THz spectroscopy and DC transport, one can see that at temperatures much lower than $T^*$ and $T_N$ an enhancement of the lowest frequency spectral weight occurs.  This is consistent with the appearance of a very narrow Drude-like peak in the low frequency conductivity.  This peak, which becomes especially pronounced at temperatures below 2.5K, is associated with the formation of a heavy fermion state.   Within the context of the Drude model, the imaginary part of the conductivity will be greatly suppressed at low frequency when the narrow $\omega = 0$ peak forms.  In contrast, at temperatures well above $T^*$ and $T_N$, the imaginary part of the complex conductivity is flat and small in comparison to the real part.

Using the DC data, the low temperature THz conductivity can be roughly fit to a Drude model using two zero-frequency oscillators and a high frequency dielectric constant $\epsilon_\infty$, which accounts for high frequency contributions outside of the measured spectral range \cite{Kuzmenko05a}.  These fits are highly constrained by the use of both THz and DC data.  Fits for both the real and imaginary parts of the complex conductivity are shown as dashed lines in Fig. \ref{Cond}.  The model shows that two Drude peaks are sufficient to fit the data:  one that is large and broader than the experimental spectral range, the other narrower that accounts for the heavy fermion renormalizations at low T.    Overall the model is a reasonable parametrization of the measured spectra.  An assumption of this fitting is that there are no peaked spectral features in $\sigma_1$ between the lower end of our spectral range and DC \cite{Dressel02a,Scheffler05a}.   $A$ $posteriori$ motivation for this assumption can be found in the fact that effective masses we extract from the analysis below are in close agreement with those found via thermodynamic measurements.   Additional motivation comes from the behavior of the frequency dependent scattering rate discussed below.

An extended Drude model analysis \cite{Allen77a} can be used to characterize the data further.   In this formalism, a frequency dependent mass $m^*(\omega)/m_b$ and scattering rate $1/\tau(\omega)$ are extracted from the measured optical constants by the relations
\begin{equation}
\label{mass}
\frac{m^*(\omega)}{m_{b}}=-\frac{\omega_{p}^2}{4\pi\omega}Im \left[\frac{1}{\sigma(\omega)} \right]
\end{equation}

\begin{equation}
\label{scatrate}
\frac{1}{\tau(\omega)}=\frac{\omega_{p}^2}{4\pi}Re \left[\frac{1}{\sigma(\omega)}\right]
\end{equation}

\noindent where $\omega_{p}$ is the plasma frequency that parametrizes the total Drude spectral weight (proportional to $\omega_{p}^2$) at high temperature, and $m_b$ is the band mass.  We performed Fourier transform infrared spectroscopy (FTS) in a transmission geometry up to 7000 cm$^{-1}$ at room temperature that allows us to extract the full plasma frequency of the Drude term $\omega_{p}=2\pi\cdot42600$ cm$^{-1}$ through Drude/Drude-Lorentz model fits.

\begin{figure}
\begin{center}
\includegraphics[width=1\columnwidth]{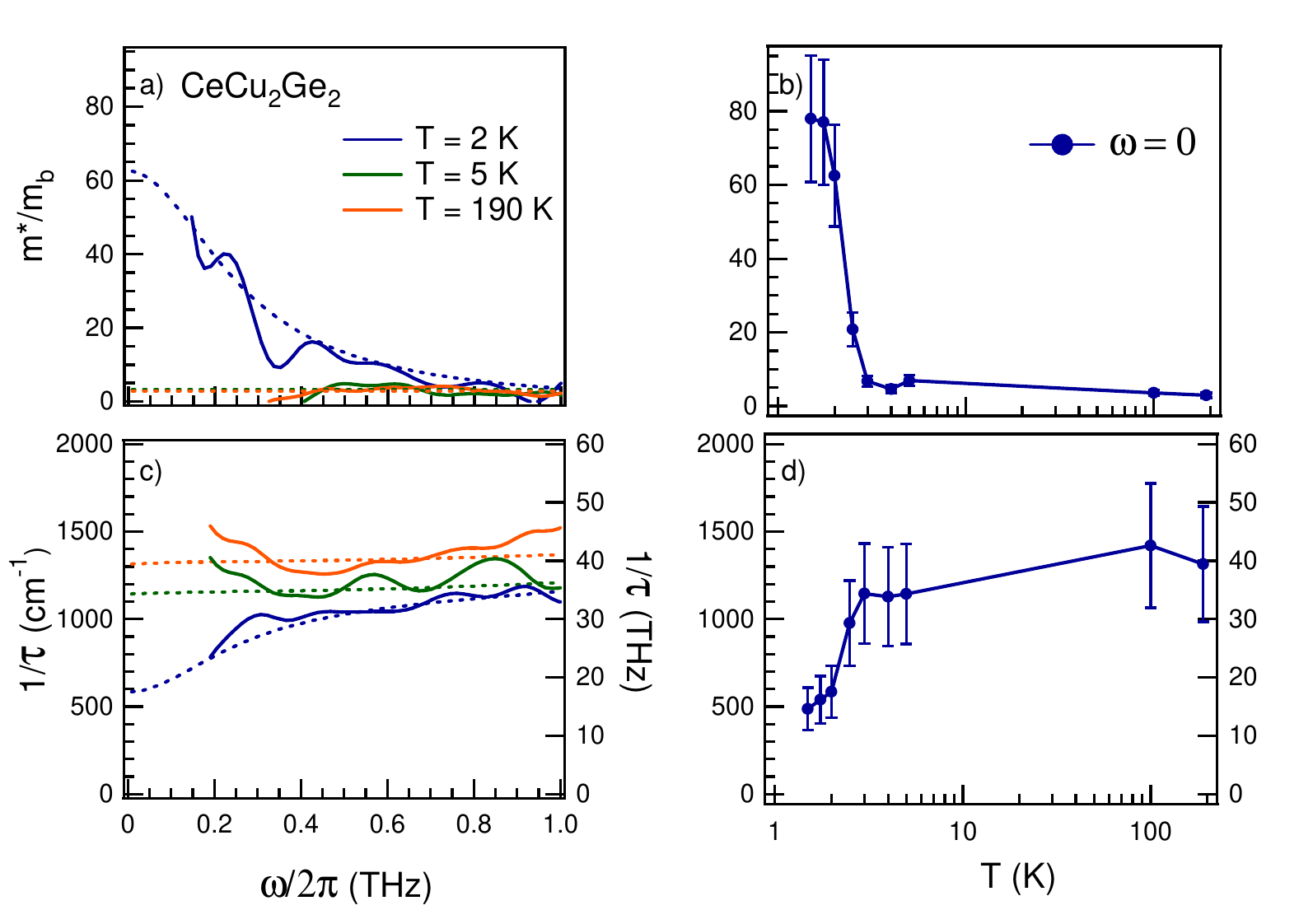}
\centering
\caption{(Color Online) a) The renormalized mass as a function of frequency derived from the extended Drude model for three temperatures.  Solid lines indicate experimental data with dashed lines showing results of a fit described in the text. b) Renormalized mass at zero frequency as a function of temperature based on the zero frequency extrapolation of the extended Drude model fits. Note the log scale on temperature axis. c) Scattering rate as a function of frequency with numerical fits shown in dashed lines for three temperatures. d) Scattering rate at zero frequency as a function of temperature.  The error bars indicate uncertainty, which is primarily due to the ambiguity in determining $\omega_p^2$ from FTS measurements.}
\label{fig:QH}
\end{center}
\end{figure}

In Fig. \ref{fig:QH}a we show the renormalized mass as a function of frequency between 150GHz and 1 THz for three temperatures using the expressions Eq. \ref{mass} and \ref{scatrate}.   At high temperatures, both the mass and scattering rate are nearly frequency independent as expected for a broad featureless conductivity.   For temperatures below the range of $T^*$ and $T_N$, a strong temperature and frequency dependent mass enhancement is observed below 300 GHz.  This suggests the presence of heavy quasiparticles at low temperatures and frequencies.   Likewise, Fig. \ref{fig:QH}c shows the frequency dependence of the scattering rate for the same three temperatures.  As the temperature decreases below $T^*$  and $T_N$, the scattering rate is suppressed at low frequency.   Its monotonic behavior as a function of frequency gives further indication that the lower edge of our THz spectral range really does correspond to a Drude-like spectral feature peaked at $\omega =0$ and that our Drude model fits discussed above are reasonable.  The extended Drude model applied to finite frequency absorptions frequently gives unphysical behavior of the scattering rate with finite frequency peaks \cite{Mena05a}.

Well above $T^*$, the scattering rate shows no strong temperature or frequency dependence.  Extrapolations toward zero frequency using the extended Drude model inversion of the Drude model fits from Fig. \ref{Cond} are also shown as dashed lines.   Recall that these fits are strongly constrained by $both$ the DC and THz data.   The extrapolated values of $m^*(\omega)/m_b$ and $1/\tau(\omega)$ as $\omega\rightarrow{0}$ from these fits are displayed in Fig. \ref{fig:QH}b, d.  As expected at temperatures well above $T^*$  both the renormalized mass and scattering rate show little temperature dependence.  Starting near the temperatures of $T^*$  and $T_N$ the effective mass begins to show notable enhancement and continues to increase until it reaches a value of approximately 80 times the band mass at the lowest temperature.  This mass enhancement is accompanied by a drop in the scattering rate, effective at the same temperature.  The scattering rate at the lowest temperatures is nearly a third of its value at room temperature.  Note also, that the mass enhancement is very similar to the scale inferred from specific heat measurements \cite{deBoer87a}.

In keeping with the usual heavy fermion phenomenology \cite{Millis87a}, despite the huge renormalization of $m^*$, the $\omega \rightarrow 0$ limit of the conductivity only suffers modest effects (approximately a factor of 2 reduction in the experimental range).    This is because the effective scattering rate $1/\tau^*$ that enters the DC conductivity $\sigma_{DC} = \frac{Ne^2 \tau^*} {m^*} $ is related to the $\omega \rightarrow 0$ scattering rate of Eq. \ref{scatrate} by the relation $1/\tau^* = 1/ \tau \frac{m_b}{m^*}$ and hence includes both lifetime and mass renormalization effects.   The two factors of $m^*$ cancel leaving  $\sigma_{DC}$ essentially unaffected by the mass enhancement.  In the present case the changes that do exist are presumably due to the development of the spin-density wave state and the reduction of the phase space for scattering that results from it.

In summary, we have applied timed-domain THz spectroscopy to the Kondo-lattice antiferromagnet CeCu$_2$Ge$_2$.   At temperatures of order the magnetic transition temperature a substantial enhancement in the low frequency conductivity is found.   Within an extended Drude-model analysis, this enhancement can be interpreted as a strongly frequency dependent mass and scattering rate, where the mass is enhanced and scattering rate suppressed at low frequency.   In the low frequency limit, this mass has a value very close to that inferred from thermodynamic measurements.  The agreement between parameter values derived in dynamic and thermodynamic experiments is a beautiful confirmation of the quasiparticle concept as applied to these material systems.   Due to their close proximity in temperature, it is unclear from our measurements (and others) whether the onset of the mass enhancement in this system should be more closely identified with $T_N$ or $T^*$.   Irrespective of this, it is clear that the enhancements to $m^*$ are occurring most strongly below 3K, which is well into the antiferromagnetic state.  Although the conventional wisdom is that the heavy mass state should develop at $T^*$, we point out that changes in the temperature dependence of $m^*$ may even be expected at $T_N$ as the Kondo interaction will be altered when the spectrum of local moment fluctuations changes with the occurrence of the ordered magnetic state and the formation of spin-waves.

Work at JHU was supported by the Gordon and Betty Moore Foundation and DOE grant for The Institute of Quantum Matter at JHU DE-FG02-08ER46544.   Work at UIUC was supported by the Center for Emergent Superconductivity, an Energy Frontier Research Center funded by the U.S. Department of Energy, Office of Science, Office of Basic Energy Sciences under Award Number DE-AC0298CH1088.  The authors would like to thank Dirk van der Marel for helpful conversations.

\bibliography{HeavyFermions}

\end{document}